\documentclass[12pt]{article}

\usepackage{graphicx} 

\usepackage{scicite}
\usepackage{times}
\usepackage{amsmath}
\usepackage{amssymb}
\usepackage{graphicx}    
\usepackage{stfloats}
\usepackage{placeins}

\newenvironment{sciabstract}{%
\begin{quote} \bf}
{\end{quote}}

\newcounter{lastnote}

\title{Wide-field fluorescence lifetime imaging of neuron spiking and sub-threshold activity \textit{in vivo}}

\author
{Adam J. Bowman$^{1\ast}$, Cheng Huang$^{2,3}$, Mark J. Schnitzer$^{2,4,5}$, Mark A. Kasevich$^{1\ast}$\\\\
\normalsize{$^{1}$Physics Department, Stanford University; 382 Via Pueblo Mall, Stanford, California 94305, USA.}\\
\normalsize{$^{2}$James H. Clark Center, Stanford University; 318 Campus Dr., Stanford, CA 94305, USA.}\\
\normalsize{$^{3}$Present Address: Department of Neuroscience, Washington University School of Medicine}\\ 
\normalsize{St. Louis, MO 63110, USA}\\
\normalsize{$^{4}$CNC Program, Stanford University; Stanford, CA, USA.}\\
\normalsize{$^{5}$Howard Hughes Medical Institute, Stanford University; Stanford, CA, USA.}\\
\\
\normalsize{$^\ast$abowman2@stanford.edu,  kasevich@stanford.edu}
}

\date{}

\begin{document} 


\baselineskip24pt


\maketitle 


\begin{sciabstract}
  The development of voltage-sensitive fluorescent probes suggests fluorescence lifetime as a promising readout for electrical activity in biological systems. Existing approaches fail to achieve the speed and sensitivity required for voltage imaging in neuroscience applications. Here we demonstrate that wide-field electro-optic fluorescence lifetime imaging (EO-FLIM) allows lifetime imaging at kHz frame acquisition rates, spatially resolving action potential propagation and sub-threshold neural activity in live adult \textit{Drosophila}. Lifetime resolutions of $<\:$5 ps at 1 kHz were achieved for single cell voltage recordings. Lifetime readout is limited by photon shot noise and the method provides strong rejection of motion artifacts and technical noise sources. Recordings revealed local transmembrane depolarizations, two types of spikes with distinct fluorescence lifetimes, and phase locking of spikes to an external mechanical stimulus.
\end{sciabstract}

\paragraph*{Introduction}
\noindent Recording the electrical activity of neurons at high spatial and temporal resolution is central to understanding brain function. Fluorescent probes of calcium activity enable optical studies of large neuron populations \textit{in vivo} \cite{Kim2022FluorescenceDynamics,Ahrens2013Whole-brainMicroscopy}.  However, the response time of calcium indicators is much slower than the underlying electrical signals. Fluorescent voltage sensors are a complementary approach, providing direct readout of neuron membrane potential with the capability to resolve action potentials. While voltage probes have rapidly developed with a variety of genetically-encoded \cite{Kannan2022Dual-polarityTypes,Tian2021All-opticalVivo,Piatkevich2019PopulationMice,Abdelfattah2019BrightImaging,Gong2015High-speedSensor} and chemical dyes \cite{Liu2020ElectrophysiologyIndicators} in use, there remain significant challenges to their application \textit{in vivo} due to low sensitivity, rapid photo-bleaching, and motion artifacts. To achieve high speed recording and sufficient signal to noise, most voltage probes use fluorescence intensity to read out an underlying sensing mechanism based on absorption, F\"{o}rster resonance energy transfer (FRET), or quenching. 

We applied an alternative strategy for detecting fast probe dynamics based on lifetime imaging \cite{Datta2020FluorescenceApplications}. Voltage sensors employing FRET and quenching mechanisms modulate the probe's non-radiative decay rate, intrinsically connecting fluorescence intensity with nanosecond excited state lifetime. Lifetime is a promising readout for voltage imaging, especially due to its capability to provide an absolute indication of membrane potential \cite{Brinks2015Two-PhotonVoltage}. Recent results have validated fluorescence lifetime for static measurements of membrane potential \textit{in vitro} \cite{Lazzari-Dean2019OpticalImaging}. However, existing lifetime detectors, e.g. single photon counters or cameras with modulated pixels \cite{Raspe}, fall short of the requirements for detecting fast dynamics and imaging neurons \textit{in vivo}, either due to their limited photon throughput or prohibitively high noise. 

\paragraph*{EO-FLIM Technique}
We demonstrate a new approach using EO-FLIM, an all-optical technique for lifetime imaging based on nanosecond gating of a wide-field image \cite{Bowman2021ResonantResolution,Bowman2019Electro-OpticMicroscopy}. EO-FLIM allows efficient photon collection and is compatible with detection on high-speed, low-noise scientific cameras. With this method, we achieved lifetime imaging of action potentials \textit{in vivo}.

EO-FLIM enables significant suppression of intensity artifacts, allowing robust imaging in the presence of tissue motion and fluctuations in illumination intensity. Such artifacts are ubiquitous in recordings of neural activity from awake, behaving animals \cite{Marshall2016Cell-Type-SpecificMice,Creamer2022CorrectingImaging}. This has two consequences: 1) it enables faithful recording of sub-threshold voltage waveforms and 2) it improves the signal-to-noise ratio by suppressing high-frequency intensity noise. These follow from the fact that in EO-FLIM, lifetime is estimated from the ratio of a pair of simultaneously recorded intensity channels deriving from a common optical source. In conventional approaches, corrections for intensity noise involve ratios of measurements which are non-simultaneous and easily corrupted by high-frequency noise or slower motion artifacts. Typically optical sensors of neuron activity are reported by $\Delta F/F$, referencing fast intensity changes ($\Delta F$) to a non-simultaneous, average fluorescence baseline ($F$). Fluorescence lifetime is capable of reading out an intensity-optimized sensor with improved temporal noise performance and long term stability without sacrificing acquisition speed.

We implemented our approach through incorporation of a Pockels cell into the fluorescence detection path of a standard epifluorescence microscope (Figs. 1(A), S1, and Methods). The Pockels cell design was optimized for resonant drive and wide-field imaging, incorporating thermal control and transverse crystal geometry to cancel off-axis birefringence. A high voltage modulation was applied to the Pockels cell using a resonant transformer (Figs. S2 and S3), resulting in a fast polarization rotation which was synchronous with the excitation pulses from a 100 picosecond laser source. Fluorescence from the sample was first polarized, then polarization modulated by the Pockels cell, and finally split by a polarizing beamsplitter into two wide-field images on a scientific CMOS camera corresponding to gated (G) and ungated (U) intensity. These two images encoded nanosecond time information in their intensity ratio.  Since gating was performed with a beamsplitter, it was possible to capture the entire fluorescence decay with photon efficiency limited by optical coatings. Here, we modulated one input polarization and discarded half of the available signal on a first polarizer.  This can be avoided in the future through addition of a second beam path \cite{Bowman2021ResonantResolution}. The fluorescence decay at each pixel was convolved with the temporal gating function of the Pockels cell and then sampled at a single modulation phase relative to the excitation laser [Fig. 1(B)]. Each image thus provided a lifetime estimate for every pixel in parallel at the framerate of the scientific camera. When imaging genetically targeted neurons \textit{in vivo}, this allowed for 1 kHz frame rate recordings with a lifetime sensitivity of $2.53 \pm 0.48$ ps with $0.7-1.4\times10^7$ detected photons per frame [Fig. 1(C)], representing a significant improvement in throughput over previous FLIM recordings of cellular dynamics. EO-FLIM approaches fundamental sensitivity limits for estimating lifetimes between 1 and 4 nanoseconds (Fig. S2). 

\paragraph*{Results}
EO-FLIM was used to image a genetically encoded voltage indicator (GEVI) in \textit{Drosophila melanogaster}. We expressed pAce positive polarity GEVI in a subtype of fly mushroom body output neuron (MBON), MBON-$\gamma$1pedc$>\alpha\beta$. pAce works via FRET and is a fusion of the bright fluorescent protein mNeonGreen with a voltage-sensitive opsin from \textit{Acetabularia} \cite{Kannan2022Dual-polarityTypes}.  We surgically prepared \textit{Drosophila} before imaging to provide optical access to the brain \cite{Sinha2013High-speedActivity,Huang2018Long-termFlies}. Action potentials and sub-threshold dynamics were readily resolved with action potentials corresponding to a 20-50 picosecond shift in the fluorescent lifetime [Fig. 1(D,E)]. Average spike readouts in lifetime and intensity for different neuron sub-regions were in strong agreement both in their relative timing and amplitudes [Fig. 1(F,G)].

The donor fluorescence lifetime of a FRET GEVI depends on radiative decay rate $k_\mathrm{f}$ and the voltage sensitive non-radiative decay rate $k_{\mathrm{nr}}(V)$ associated with FRET as $\tau = \frac{1}{k_\mathrm{f}+k_{\mathrm{nr}}(V)}$ \cite{Brinks2015Two-PhotonVoltage,Clegg2009ForsterDone}.  In pAce, the Ace opsin acts as acceptor and provides voltage sensitivity. The donor's fluorescence intensity is directly quenched by FRET giving a signal $\Delta F \propto \sigma \Delta k_{\mathrm{nr}}$ where $\sigma$ is the donor excitation cross section. For an ideal FRET process, one expects to find $\Delta\tau/\tau =\Delta F/F $. pAce gave a linear but attenuated lifetime response of $(0.70 \pm 0.07 )$$ \Delta F/F $ (Fig. S4). This may indicate components of the GEVI response in intensity, for example modulation of cross section $\sigma$, that did not affect lifetime readout.

Wide-field lifetime imaging correlates neuron activity with spatial structure. The point of action potential initiation in the axon is resolved along with bi-directional propagation along the axon and backwards towards the dendrite and soma \cite{Huang2022DopamineDynamics}. Action potentials were attenuated and broadened in the dendrites and soma [Figs. 1(F,G), S5, and S6]. Spike propagation is shown in Movies S1-S3 with still frames from Movie S1 displayed in Fig. 1(H), generated by spike triggered averaging over $\sim 300$ spikes and interpolating between frames. We also observed individual spikes and spike propagation in real time without temporal averaging (Movies S4 and S5). Comparison of recordings from multiple neuronal sub-regions revealed local depolarizations in the axons which fail to initiate action potentials across the entire cell. These were not resolved in intensity readout but are clearly visible in lifetime recordings (Figs. S5 and S6). Finally, by applying a 10 frame moving average, the spatial distribution of slower sub-threshold voltage signals could also be studied. Sub-threshold signals were often strongest and localized in the dendrites (Movies S6 and S7). Still frames from these movies are shown in Fig. 1(I) and Figs. S7 and S8.

By measuring a ratio of simultaneous intensities, EO-FLIM removes noise sources that are common mode to both the modulated channels. In Fig. 2 we show example recordings of MBON-$\gamma$1pedc$>\alpha/\beta$ neurons from six flies comparing intensity $\Delta F/F$ and lifetime readouts. In all recordings lifetime detection enhanced the signal-to-noise ratio (SNR) for action potential detection by $\sim 2$x. This SNR was quantified by comparing spike amplitude to the high frequency noise floor. We also analyzed traces using the spike detection fidelity $d'$, a discriminability index that quantifies spike detection by comparing the statistical distributions of spike amplitudes and background noise fluctuations \cite{Wilt2013PhotonTiming}. Lifetime detection improved $d'$ by 1.5 - 2.4 times in Fig. 2(A-E). 

Noise power was also compared across temporal frequencies, demonstrating broad suppression of intensity noise in Fig. 2(A-E) and showing that EO-FLIM approaches the photon shot noise limit. To allow a direct comparison of noise power spectrum, the responsivity of spikes in intensity and lifetime channels were normalized. For flies not displaying much motion [Fig. 2(A,B)], lifetime primarily reduced technical noise at high frequencies that resulted from the excitation laser (4-7 dB). Even in these well-behaved examples lifetime readout resulted in improved SNR and detection fidelity $d'$. 

Lifetime recordings also improved long-term stability in the voltage readout. Intensity-based voltage imaging often displays strong motion artifacts which degrade stability. In \textit{Drosophila}, these artifacts result from movements such as extension of the proboscis. For flies displaying motion [Figs. 2(C-E) and S9], lifetime readout suppressed artifacts at low frequencies by up to 9 dB compared to intensity readout (here sub-threshold waveforms may only be resolved in lifetime). Fig. 3 quantifies trace stability by the spike amplitude distribution and the uniformity of threshold voltage level at action potential locations. Histograms of mean normalized spike heights and mean normalized sub-threshold level are plotted for each spike, showing that lifetime improves spike uniformity by up to 2.5 times and threshold uniformity by up to 5.8 times. 

With the improved readout stability afforded by lifetime recording, we observed two spike amplitudes in Fig. 2(F-L). The small amplitude spikes occured on top of sub-threshold voltage plateaus while large amplitude spikes were observed in bursts that were independent of sub-threshold voltage level. In this sample, the fly was mechanically stimulated near a resonance of the microscope stage with sound waves. Fig. 2(I,J) plot histograms of spike heights in both intensity and lifetime. The two spike populations were clearly distinguished using lifetime readout but were not separable using intensity. Here the large spikes showed enhanced lifetime responsivity (1.34 $\Delta F/F$, compared to 0.68 $\Delta F/F$ for smaller spikes). The difference in responsivity may indicate kinetic differences in the GEVI response to different action potential waveforms \cite{Kannan2022Dual-polarityTypes}. Spike triggered averages (Movies S8 and S9) showed that the L spikes originated diffusely across the image while the S spikes initiated in a local area of the axon as in Movies S1 and S2. The L spikes may be associated with out of focus neurons displaying off-target GEVI expression, and they were also synchronous with spiking of the targeted neuron as demonstrated in Fig. 2(F-H). The L spikes displayed strong phase locking in response to mechanical stimulus, while the S spikes did not. A histogram of phases is displayed in Fig. 2(K) where phase is determined by each spike's location on a trace that has been filtered at the stimulus frequency. Narrow-band locking response is demonstrated by Fig. 2(L). Similar large amplitude spikes were also observed during the mechanical stimulus sweep shown in Fig. 4. 

To further demonstrate the noise rejection capabilities of lifetime detection, we placed the fly on a piezoelectric stage to provide mechanical shaking in the XY plane. This direct mechanical stimulus resulted in high levels of intensity noise that obscured neuron activity, but lifetime readout suppressed this noise by up to 21 dB in Fig. 4(A,B). Using this direct stimulus we could observe phase locked spiking behavior from 30 to $>$100 Hz in Figs. 4 and S10. In these figures, phase locking was observed through increased spectral power at the stimulus frequency in a spectral waterfall plot as the excitation frequency was swept. This phase locking may also be visualized using the autocorrelation of the lifetime trace [Fig. 4(C-D)]. These observations are consistent with previous studies on mechanical and auditory effects in \textit{Drosophila} that identify a broad auditory response across the central brain \cite{Pacheco2020AuditoryDrosophila} and responses to substrate vibrations \cite{Hehlert2021DrosophilaTransduction,Zhang2013SoundLarvae}. 

\paragraph*{Discussion}
EO-FLIM may be applied to both existing lifetime-sensitive probes and to donor readout of FRET-based biosensors. Standard FRET sensors are read out by the ratio of optical intensities in spectrally separated donor and acceptor channels \cite{Akemann2012ImagingProtein}, requiring an acceptor molecule with high quantum yield. Two-color readout frequently limits detection sensitivity due to spectral cross-talk and also prevents probe multiplexing \cite{Grant2008MultiplexedCells}. Lifetime measurement removes these limitations and allows quantitative FRET measurements using only the donor channel. In voltage imaging, lifetime will enable improved measurement of FRET-opsin \cite{Kannan2022Dual-polarityTypes,Gong2015High-speedSensor}, hybrid FRET \cite{Abdelfattah2019BrightImaging}, and dye indicators \cite{Liu2020ElectrophysiologyIndicators}. We also anticipate application to imaging calcium \cite{vanderLinden2021ACalcium}, neurotransmitters \cite{Ma2022FluorescenceAnimals,Sistemich2023NearNanotubes}, and cyclic AMP\cite{Klarenbeek2015Fourth-GenerationAffinity}.

Use of GEVIs \textit{in vivo} is often accompanied by a large fluorescence background that results from protein expression outside the cellular membrane \cite{Platisa2018GeneticallyYet} or leaky gene expression from non-targeted cells \cite{Aso2014TheLearning}. This background signal had a different fluorescence lifetime and photobleaches at a different rate, resulting in slow drifts of the measured lifetime traces as shown in Fig. S11 (we expect \textit{in vitro} studies will not be affected by such backgrounds). To measure absolute voltage, signal and background populations would need to be unmixed by discriminating between two closely spaced exponential decays. For this reason, we focus here on the improved stability and noise performance afforded by lifetime measurement rather than absolute quantification. In our current implementation, we measured the population weighted average lifetime by acquiring images at a single modulation phase. In the future, multiple modulation phases can be combined to unmix lifetime components and improve absolute measurement. 

We have shown that EO-FLIM enables fluorescence lifetime imaging of neuron activity \textit{in vivo}, overcoming the throughput and sensitivity limitations of existing FLIM techniques. We expect straightforward application to other systems, including mammalian brains which feature both larger neurons and action potentials compared to \textit{Drosophila}\cite{Kannan2022Dual-polarityTypes,Gong2015High-speedSensor}. Voltage imaging in neuroscience is one example application, but membrane potential is also broadly interesting throughout biology from bacteria \cite{Prindle2015IonCommunities,Kralj2011ElectricalProtein} and plants \cite{Serre2021AFB1Root} to cardiac \cite{Acker2020RecentHearts,Lee2019InModels,Wikswo1995VirtualStimulation} and muscle tissue \cite{Hashemi2019Rhodopsin}. The ability of EO-FLIM to strongly reject motion noise \textit{in vivo} is relevant for brain and cardiac imaging applications where it is challenging to faithfully distinguish voltage dynamics from motion and hemodynamic artifacts \cite{Marshall2016Cell-Type-SpecificMice,Lee2019InModels, Acker2020RecentHearts, Creamer2022CorrectingImaging}. It may become possible to perform voltage imaging during natural movements such as insect flight, or while tracking a freely moving organism \cite{Nguyen2016Whole-brainElegans}.  Further, recent advances in GEVI probes have enabled voltage imaging of populations of neurons \cite{Piatkevich2019PopulationMice,Kannan2022Dual-polarityTypes}. Lifetime imaging will establish accurate long-term readout of sub-threshold activity across a neural circuit, allowing functional connectivity mapping where spike activity may be correlated with sub-threshold modulation of downstream neurons. Finally, using lifetime detection in combination with optogenetic tools \cite{Adam2019VoltageDynamics,Tian2021All-opticalVivo}, it will be possible to improve techniques for targeted optical activation and control \cite{Bergs2022All-opticalAnimals} of neuron membrane potential.

\clearpage


\bibliographystyle{Science}
\clearpage

\subsection*{Acknowledgements}
\noindent\textbf{Funding:} We acknowledge funding from the Gordon and Betty Moore Foundation, the U.S. Department of Energy, Office of Science, Office of Biological and Environmental Research under Award Number DE-SC0021976, NIH grant U01NS120822 (M.J.S. and Ganesh Vasan), and NSF NeuroNex grant DBI-1707261 (M.J.S. and K. Deisseroth). A.B. acknowledges support from the NSF Graduate Research Fellowship under grant 1656518 and the Stanford Graduate Fellowship.\\

\noindent\textbf{Author Contributions:} A.B. developed the microscope. C.H. prepared \textit{Drosophila} for imaging. A.B. and C.H. performed the experiments. A.B. and M.K. analyzed data and wrote the manuscript. All authors contributed to experiment conception and manuscript revision.\\

\noindent\textbf{Competing Interests:} A.B. and M.K. are inventors on  PCT/US2019/062640, US17/153438, and US17/898093. \\

\noindent\textbf{Data Availability:} All data are available in the manuscript, the supplementary material, or deposited at Dryad \cite{Dryad}. Code is available at Zenodo \cite{Zenodo}. 
\\
\subsubsection*{Supplementary Materials}
Materials and Methods\\
Figs. S1 to S11\\
Movies S1 to S9

\clearpage
\begin{figure*}[h!]
\centering
\vspace{-1in}
  \includegraphics[width=1\textwidth]{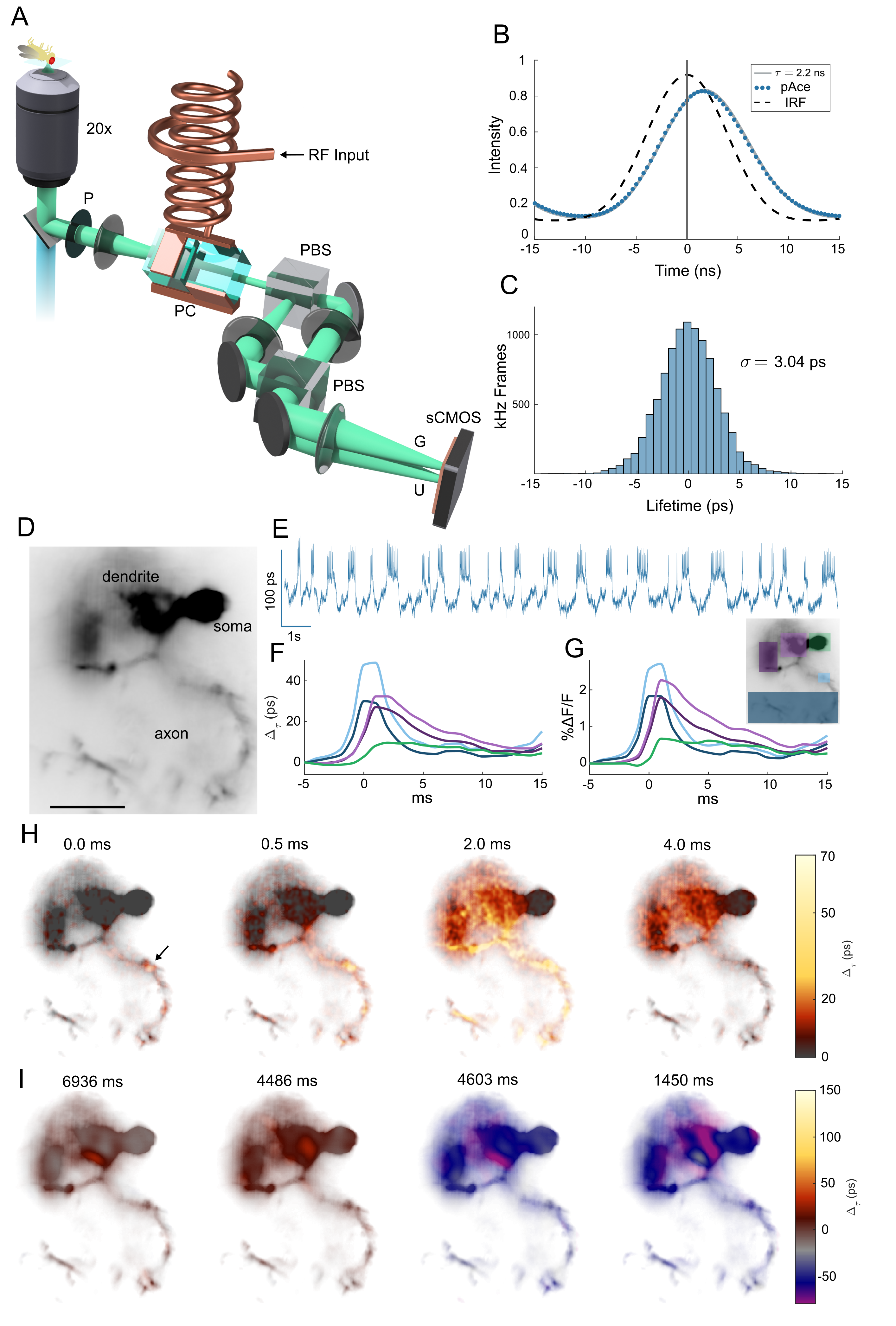} 
  \vspace{-5.0mm}
\end{figure*}

\clearpage

\noindent {\bf Fig. 1.} \textbf{Lifetime imaging of action potentials} (\textbf{A}) Schematic of EO-FLIM microscope. Wide-field fluorescence images were modulated by a Pockels cell (PC) placed between crossed polarizers (P and PBS) and driven at 20 MHz by a high voltage resonant transformer. Two spatially offset output images were simultaneously captured after a second polarizing beamsplitter (PBS) on a sCMOS camera, corresponding to gated (G) and ungated (U) intensities. (\textbf{B}) Instrument response function (IRF) and fluorescence traces for the U channel were measured by varying the Pockels cell drive phase relative to the excitation laser. The pAce GEVI was fit to 2.2 ns lifetime. For kilohertz imaging, a single optimal phase point was captured (vertical line at 0 ns delay) and the G/U image intensity ratio was converted to a lifetime estimate (see also Fig. S1). (\textbf{C}) Histogram of measurements (highpass filtered) obtained at 1 kHz for a single neuron \textit{in vivo} demonstrate a lifetime sensitivity of 3 ps (full trace in Fig. 2(A)) (\textbf{D}) Wide-field image of a neuron with structures indicated (scalebar 25 $\mu$m) (\textbf{E}) Whole cell lifetime trace resolves action potentials and sub-threshold transitions (\textbf{F}, \textbf{G}) Average spike shape is plotted in intensity and lifetime from color-coded regions (\textbf{H}) Frames from an interpolated lifetime movie demonstrate spike propagation, averaging the signal from $\sim$300 individual spikes. The point of  initiation is indicated by the arrow, and bidirectional propagation was observed both along the axon and backwards towards soma and dendrites (see Movies S1-S3). Spike propagation was also imaged directly without averaging in Movies S4 and S5. (\textbf{I}) Applying a 10 frame moving average allowed sub-threshold signals to be localized to neuron structures in Movies S6 and S7. Example frames demonstrate localization in the dendrite for both positive and negative sub-threshold signals.

\clearpage
\begin{figure*}[h!]
\centering
\hspace*{-0.5in}
  \includegraphics[width=1.15\textwidth]{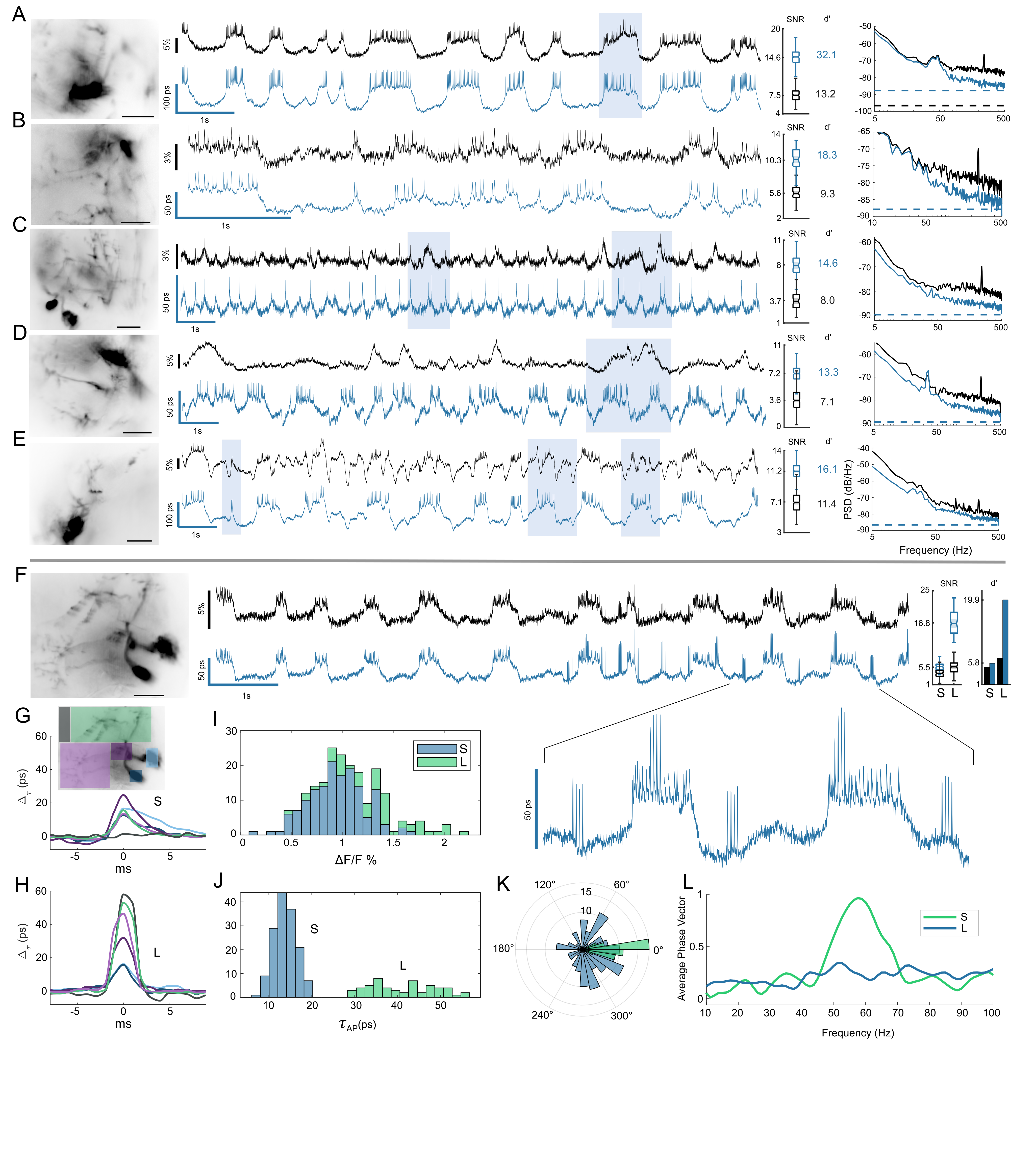} 
  \vspace{-5.0mm}
\end{figure*}

\clearpage

\noindent {\bf Fig. 2.} \textbf{Lifetime suppresses intensity noise and improves fidelity of sub-threshold recording.} Six example MBON neurons are shown comparing lifetime (blue) to $\Delta F/F$ intensity recordings (black). Recordings were obtained by averaging over high-resolution images shown at left (scalebar 25 $\mu$m). Shaded boxes highlight some notable regions of the traces for improved lifetime readout. For each example, the distributions of spike SNR are compared for intensity and lifetime, with calculated spike detection fidelity $d'$ indicated. (\textbf{A},\textbf{B}) Two examples of flies without motion demonstrate improvement of technical noise floor at high frequencies by up to 7 dB. The noise power spectra for the traces are compared at right, with dotted lines indicating the photon shot noise limits (\textbf{C}-\textbf{E}) Three examples of flies having low-frequency noise associated with motion artifacts. Lifetime improves noise power spectrum across temporal frequencies, rejecting intensity noise by up to 9 dB at low frequencies. See also further analysis in Fig. 3. (\textbf{F}-\textbf{J}) Lifetime provided an improved readout of two spike amplitudes in response to mechanical stimulus at 60 Hz. Large (L) spikes showed an enhanced lifetime responsivity and tripled detection SNR and $d'$ over the small (S) spikes. L spikes occured independent of sub-threshold waveform level but synchronized with spiking on plateaus in the inset. (\textbf{G},\textbf{H}) Average spike waveforms for color-coded regions. The point of initiation for  S spikes was a central region of the axon (consistent with Movies S1 and S2), while L spikes were diffuse and associated with background fluorescence. L spikes also correspond to local spikes in the dendrite and soma in (H). L spike background component possibly resulted from out of focus neurons (Movies S8 and S9). (\textbf{I},\textbf{J}) Histograms of action potential amplitudes are compared. In intensity the L and S populations were not resolved and strongly overlap, but they were clearly separated in lifetime. Using lifetime to identify the spikes, the intensity histogram (I) is shaded with two colors to show overlapping populations. (\textbf{K}) A polar histogram demonstrates strong phase locking of the L spikes to mechanical stimulus using a bandpass filtered lifetime trace as phase reference. S spikes do not show phase locking. (\textbf{L}) The average phase vector length $\sum_i \cos(\theta_i)/N_{\mathrm{AP}}$ is plotted vs. bandpass center frequency to show narrow-band locking response.

\clearpage
\begin{figure*}[h!]
\centering
  \includegraphics[width=0.8\textwidth]{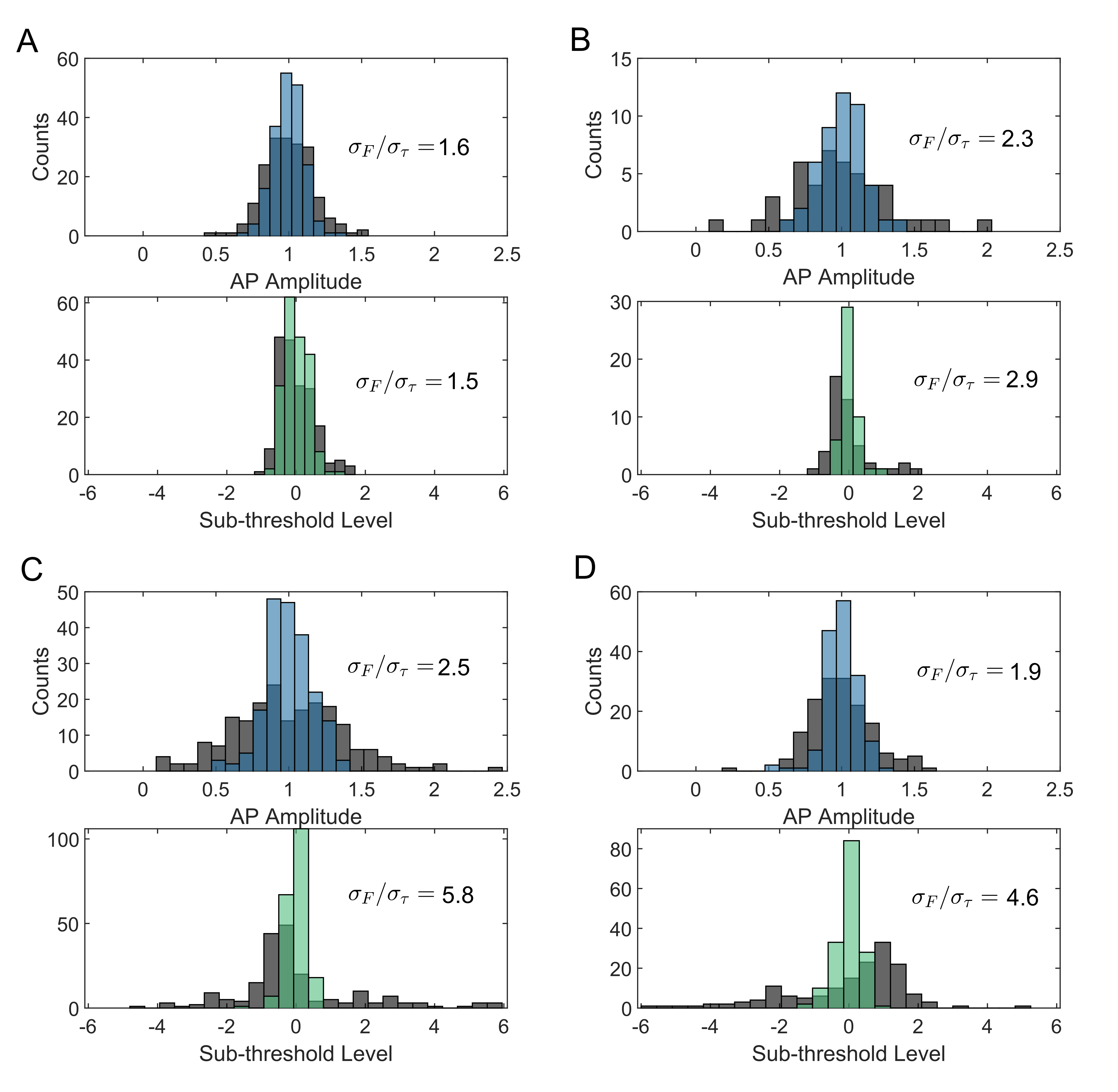} 
  \vspace{-5.0mm}
  \end{figure*}

\clearpage

\noindent {\bf Fig. 3.} \textbf{Lifetime improves uniformity in action potential amplitude and threshold level.} Histograms of action potential amplitudes (lifetime in blue) and action potential levels on the sub-threshold waveform (lifetime in green) are plotted for each activity trace, overlayed on the same histograms for intensity (grey).  Action potential amplitudes are normalized to the mean. Sub-threshold level is also mean normalized as $(L-\overline{L})/\overline{L}$ where $L$ is the spike's corresponding level on a low-pass filtered trace, and its distance is measured relative to the mean level of all other spikes $\overline{L}$.  A perfectly uniform threshold would thus result in L = 0 for all spikes. In each histogram, the ratio of standard deviation between intensity and lifetime readouts $\sigma_F/\sigma_\tau$ is given as a figure of merit for uniformity. Panes (A, B-D) correspond to panes (A, C-E) in Fig. 2 respectively. 

\clearpage
\begin{figure*}[h!]
\centering
\hspace*{-0.5in}
  \includegraphics[width=1.15\textwidth]{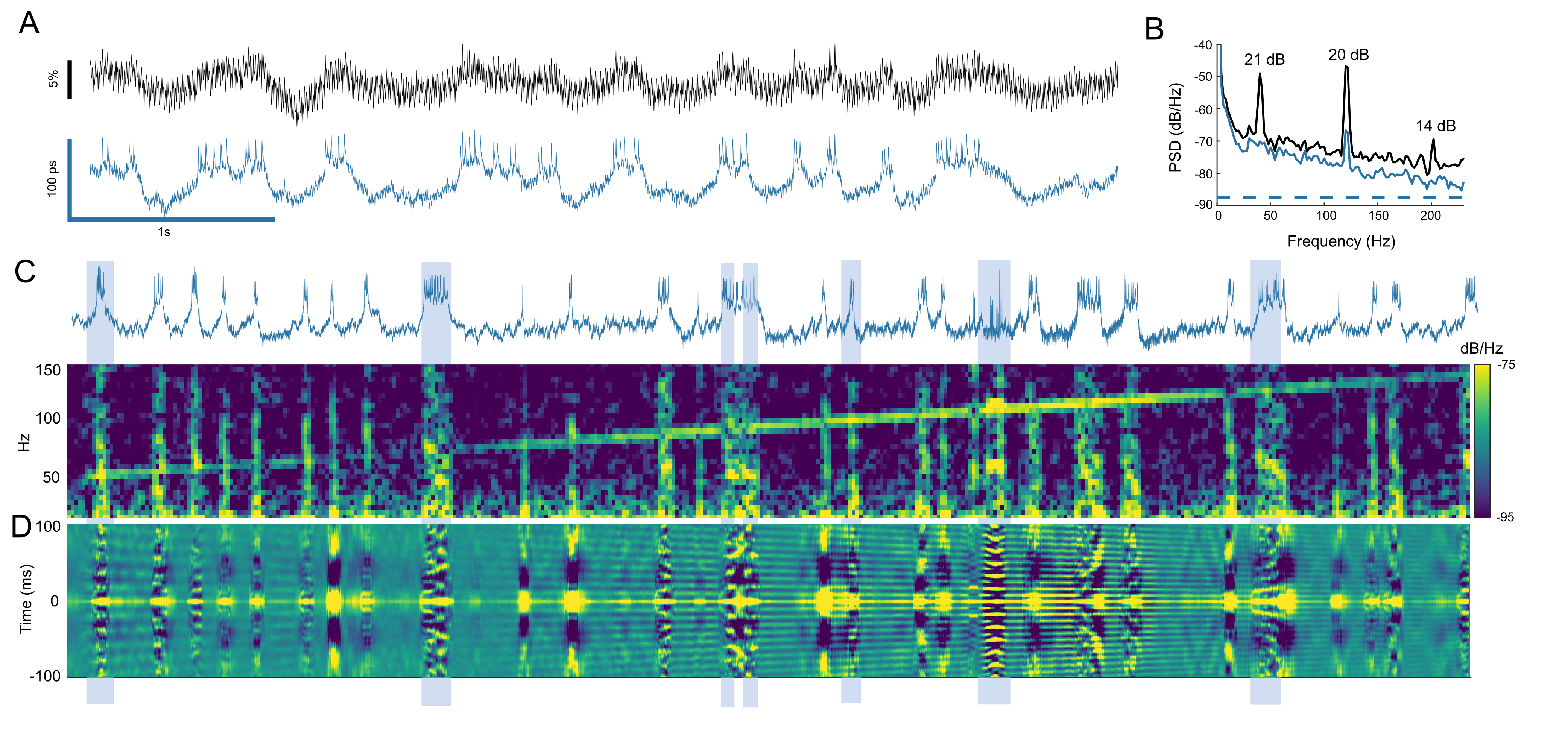} 
\end{figure*}

\clearpage

\noindent {\bf Fig. 4.} \textbf{Phase locking of spikes to direct mechanical stimulus} (\textbf{A}) Lifetime provided strong rejection of intensity noise associated with shaking the sample (40 Hz square wave, $\sim 0.8\: \mu$m peak-to-peak). (\textbf{B}) Intensity noise at the stimulus frequency and second harmonic were attenuated by 21 and 20 dB. (\textbf{C}) A spectrogram of the lifetime trace is plotted as stimulus is swept from 50 to 150 Hz. Vertical lines of activity in the spectrogram correspond to spike bursts in the lifetime trace. Mechanical cross-talk is seen as the diagonal line sweep, and phase locking appears as increased frequency content at the stimulus frequency during spike bursts. (\textbf{D}) To show phase locking visually, a sliding window autocorrelation of the lifetime trace is plotted using a 150 ms window. Phase locking may be seen by observing alignment of autocorrelation peaks during activity bursts to the peaks resulting from mechanical cross-talk signal. Examples of bursts showing phase locking are highlighted.

\newpage
\noindent\textbf{\large Materials and Methods:}

\noindent{\textbf{Optical design.}  Optics were designed around a Zeiss Axiovert S100TV microscope body. The sample is excited using a supercontinuum laser source with internal pulse picker to halve the repetition rate to 39 MHz (NKT SuperK EXW-12). The laser is first filtered with a colored glass filter to remove infrared (Thorlabs FGS900), and a FITC filter set is used for imaging (Semrock: excitation FF02-475/50, dichroic FF506-Di03, and emission FF01-540/50). The sample is imaged in epifluorescence mode using an Olympus 20X/0.95 NA XLUMPlanFl water immersion objective and excited with 1.5 mW total power ($\sim$5 W/cm$^2$ intensity at the sample). Images are acquired at 1 kHz frame rate using a Hamamatsu ORCA-Quest camera operating in standard readout mode with 0.44 electron/pixel read noise.

A first pair of relay lenses after the output port of the microscope creates an image plane near the Pockels cell crystals, and a second pair of lenses is used to relay this image onto the camera sensor. The Pockels cell is placed between polarizing beamsplitter cubes (Thorlabs PBS251) along with a liquid crystal variable waveplate that is used to zero static birefringent phase shift (Thorlabs LCC1421-A). In order to ensure equal optical path lengths and to allow independent focusing of the two output images, a third polarizing beamsplitter cube is used to combine the separated paths before the final relay lens as shown in Fig. 1(A). By placing an image plane near the Pockels cell crystals, it is easier to align the crystals such that optimal performance is obtained in the presence of thermal stress.
\\}

\noindent{\textbf{Pockels cell and RF system.}  A custom lithium tantalate Pockels cell was developed with Leysop Inc. A pair of crystals (9x9x10 mm each) is used with electric field applied transverse to the optical axis (Fig. S1). The crystals are rotated 90 degrees relative to each other in a standard configuration that also enables cancellation of off-axis birefringent phase shifts and proves optimal for wide-field imaging (compared to common DKDP crystals which have limited angular acceptance (\textit{14}) ).  Lithium tantalate is also a good choice for imaging applications due to its low birefringence. Dielectric losses dissipate power within the crystals, so they are cooled by thermally conductive aluminum nitride ceramic plates connected to a water cooling circuit. 

For an applied voltage V, the Pockels cell produces a birefringent phase shift $\delta = 2\pi (n_e^3 r_{33} - n_o^3 r_{13})VL/{d\lambda}$ where $n_o$ and $n_e$ are the ordinary and extraordinary refractive indices, $r_{33}$ and $r_{13}$ are the electro-optic coefficients, $L$ is the crystal length, and $d$ is the aperture. The voltage corresponding to a $\pi$ phase shift ($V_\pi$) for the Pockels cell used here is approximately 1.2 kV (Fig. S3). The time-dependent birefringent phase shift results in an intensity modulation through crossed polarizers given by $g(t) = \sin^2(\delta(t)/2)$ and $u(t) = \cos^2(\delta(t)/2)$ for the gated (G) and ungated (U) channels respectively. For resonant drive, this results in a gating function of the form $g(t) = \sin^2[A\pi\sin(\omega t)/2]$ where $A$ is the drive strength in units of $V_\pi$. Note that positive and negative voltage across the Pockels cell result in the same transmission, so that the optical gating waveform has twice the frequency of the drive voltage.

The Pockels cell and its resonant circuit are housed in a grounded aluminum enclosure, shown in Fig. S2. The Pockels cell acts as a capacitive load in a resonant transformer. The primary is a half turn copper strap, and the secondary comprises 11 turns of 3/16" copper tubing electrically in series with the Pockels cell electrodes (Fig. S3). This transformer achieves $\sim V_\pi$ at 19.5 MHz using 9.3 W RF power and presents an impedance-matched 50 $\Omega$ load to the drive electronics. Fine tuning of the resonant frequency is accomplished by monitoring S11 on a network analyzer (HP 4395A/87511A or NanoVNA) and adjusting a 10 pF trimmer capacitor (Sprague-Goodman SGNMNC1103) using a stepper motor.  Drive is provided by a class-A amplifier (ENI 325LA) with transmitted and reflected power monitored on a SWR meter (Youmei RS-70). The RF signal originates from the laser and is used as input to a direct digital synthesizer (Novatech 409B-AC) that produces a phase locked 19.5 MHz drive with a computer controlled phase offset. This system has proven highly stable, and it allows for continuous modulation without phase drift.
\\}

\noindent{\textbf{Sample preparation.} To express the pAce indicator in MBON-$\gamma$1pedc$>\alpha/\beta$ neurons, we crossed the MB085C-GAL4 flies with 20×UAS-pAce flies and collected their progenies for imaging experiments. To create an imaging window in the fly cuticle, we mounted flies onto a customized holder and used a laser microsurgery system based on a 193 nm wavelength excimer laser (GamLaser; EX5 ArF), as detailed in our prior work (\textit{17}, \textit{18}). In each fly, the laser microsurgery created a 150 $\mu$m diameter hole (30–40 laser pulses, delivered at 100 Hz, 36 $\mu$J per pulse as measured at the specimen plane). Immediately after surgery, we applied 1 $\mu$L of UV epoxy (NOA 68, Norland; Refractive index: 1.54; Transmission 420–1000 nm: $\sim100\%$) and cured it for 30 s to seal the cuticle opening. Then, we glued a coverslip (22 × 22 mm, No. 0, Electron Microscope Sciences) above the fly’s head to be compatible with water immersion imaging.
\\}

\noindent{\textbf{Mechanical stimulation.}  We used two methods to apply mechanical stimulation to the fly. In the first method a 5 inch speaker was placed 6 cm above the sample plane and driven to produce 85-90 dBA at the sample. This was effective at vibrating the sample stage near resonance frequencies without much in-plane motion, and resulted in observed phase locking such as shown in Figs. 2 and S10(A). For the second method, to provide a direct and tunable in-plane shaking stimulus, we used a small piezoelectric stage (Physik Instrumente P-915K238) driven with a function generator. This allowed stimulus frequency sweeps in Figs. 4 and S10(B). 
\\}

\noindent{\textbf{Analysis.}
The gated (G) and ungated (U) image pixels are registered using an affine transformation that is determined from a bright-field image of a grid micrometer slide (Electron Microscopy Sciences slide S29). In order to measure the gating functions $g(t)$ and $u(t)$ which serve as the instrument response function (IRF), a quenched fluorescein standard is used. This standard is prepared using 170 $\mu$L saturated fluorescein solution added to 1 mL of saturated potassium iodide solution which has been brought to pH 10 with potassium hydroxide. The G and U intensities measured on the camera are determined by the convolution of the IRF with the unknown lifetime $I(t,\tau)$. For a particular Pockels cell drive phase $\phi$, $G(\phi,\tau) = \int I(t,\tau)g(t-\phi/\omega) \mathrm{d}t$ and $U(\phi_0,\tau) = \int I(t,\tau)u(t-\phi/\omega)\mathrm{d}t$, where the gating functions are normalized such that $g(t)+u(t)=1$.  These convolutions were directly measured by varying the Pockels cell phase $\phi$ relative to the laser (\textit{13}). The resulting intensity curve (Figs. 1(B) and S1(D)) may be fit to determine the lifetime by sampling at multiple phase points. Alternatively, a single optimal phase $\phi$ may be chosen where intensity ratios can be used to estimate lifetime. For the selected phase, a lookup table is calculated by numerically convolving the experimentally measured IRF with single exponential decays. This table allows conversion between measured intensity ratio G/U and a lifetime estimate (Fig. S2(B-D)). Noise limits are fundamentally determined by shot noise in the G and U images, and the achievable lifetime estimation sensitivity is plotted in Fig. S2(B). When generating lifetime traces, the G/U intensity ratio, IRF, and resulting lifetime lookup table are all determined locally from the same image region of interest.

To generate lifetime movies, lifetime images are determined from pixel intensity ratios and overlayed with a transparency mask generated from the intensity image to show sample structure. Motion correction is first applied using piecewise affine transformations that are determined every 20 frames. For spike-triggered averaging, spikes locations are identified using a highpass filtered trace, then images corresponding to a time window before and after each spike are combined into a single movie that is temporally averaged over $\sim 300$ spikes referenced to the peak. Triggered movies are interpolated by generating 20 time points per millisecond through linear interpolation of each pixel trace. Due to the rolling shutter readout of the sCMOS camera, there is a small time delay between readout of different sensor regions that depends on the pixel row number. This rolling shutter delay is corrected by applying a shift to the interpolated pixel time traces using the known row readout time of the camera sensor. For non-averaged movies of sub-threshold signals, a moving average of ten frames is used to reveal slower spatial dynamics. A moving average image calculated from 1000 frames is also subtracted to remove long-term drifts. Gaussian filtering is applied to the lifetime movies to average over spatial noise. Average spike waveforms displayed in Figs. 1(F,G), 2(G,H), S5 and S6 correspond to the average spike calculated across indicated image ROIs. Here 20 points per millisecond are interpolated, and rolling shutter corrections are applied using the average row value for each ROI. In Figs. 1 and 2, piecewise cubic interpolation is used. In Figs. S5 and S6, spline interpolation is used and spike timing is determined from the peak. For comparing recordings in the different neuron sub-regions, a three frame moving average is applied.

In Fig. 2, signal-to-noise ratio (SNR) is determined by comparing spike amplitude to the RMS noise of the fluorescence baseline. Spike amplitude is determined by subtracting a 20 Hz low-pass filtered waveform from the activity traces. The RMS noise floor is determined by first replacing each spike in an interval of $\pm$3 points on each side of the spike time with the moving average. The resulting trace is then highpass filtered at 50 Hz and the standard deviation corresponds to the RMS noise value. The RMS noise distribution is shown in Fig. 1(C) for an example trace and is also used to calculate the lifetime sensitivity. Lifetime sensitivity and responsivity values are calculated using all seven neurons shown in Figs. 1 and 2. The power spectral density (PSD) is determined using Welch's method applied to the spike-removed traces described above. Because responsivities $\Delta F/F$ and $\Delta\tau/\tau$ are not equal, the lifetime traces and shot noise limits are normalized by a correction factor determined by comparing intensity and lifetime spike amplitudes. Otherwise the lifetime trace would have an artificially reduced noise level and direct comparison of power spectra would not be possible. The spike detection fidelity $d'$ is calculated from the raw data traces using a data interval of 12 ms on each side of the spike time (\textit{20}, \textit{21}). To analyze phase locking in Fig. 2(K,L), a bandpass filtered trace (5 Hz bandwidth, infinite impulse response), is used to derive a phase reference by applying a Hilbert transform. Each spike location on the original trace is then assigned a phase value using the reference. The length of the average phase vector $\sum_i \cos(\theta_i)/N_{\text{AP}}$ vs. frequency is plotted by scanning the bandpass center frequency.

Spectrograms are determined using Welch's method. In Fig. 4(C), phase locked activity is observed in the spectrogram as increased power at the stimulus frequency during spike activity bursts. By plotting the autocorrelation in Fig. 4(D) along the y-axis as a function of time (150 ms window length), it is possible to visualize the phase synchronization of spikes to the mechanical cross-talk, which is visible as a diagonal line at the stimulus frequency in the spectrogram. Phase locking is seen at either the fundamental or half-harmonic stimulus frequencies here (i.e. spiking every cycle or every other cycle). We also note that the start and stop of mechanical cross-talk signal tends to coincide with a spike burst as shown in Fig. S10(B).

No data exclusion criteria were preestablished. Sample sizes were determined from the standards of the voltage imaging field. Experimental results were biologically replicated across multiple flies ($>$30) and days of imaging. Further, technical replication was performed at each image frame throughout the recordings.

}

\renewcommand{\figurename}{\textbf{Fig.}}
\renewcommand{\thefigure}{\textbf{S\arabic{figure}}}
\setcounter{figure}{0}

\newpage
\begin{figure*}[h!]
\centering
  \includegraphics[width=1.0\textwidth]{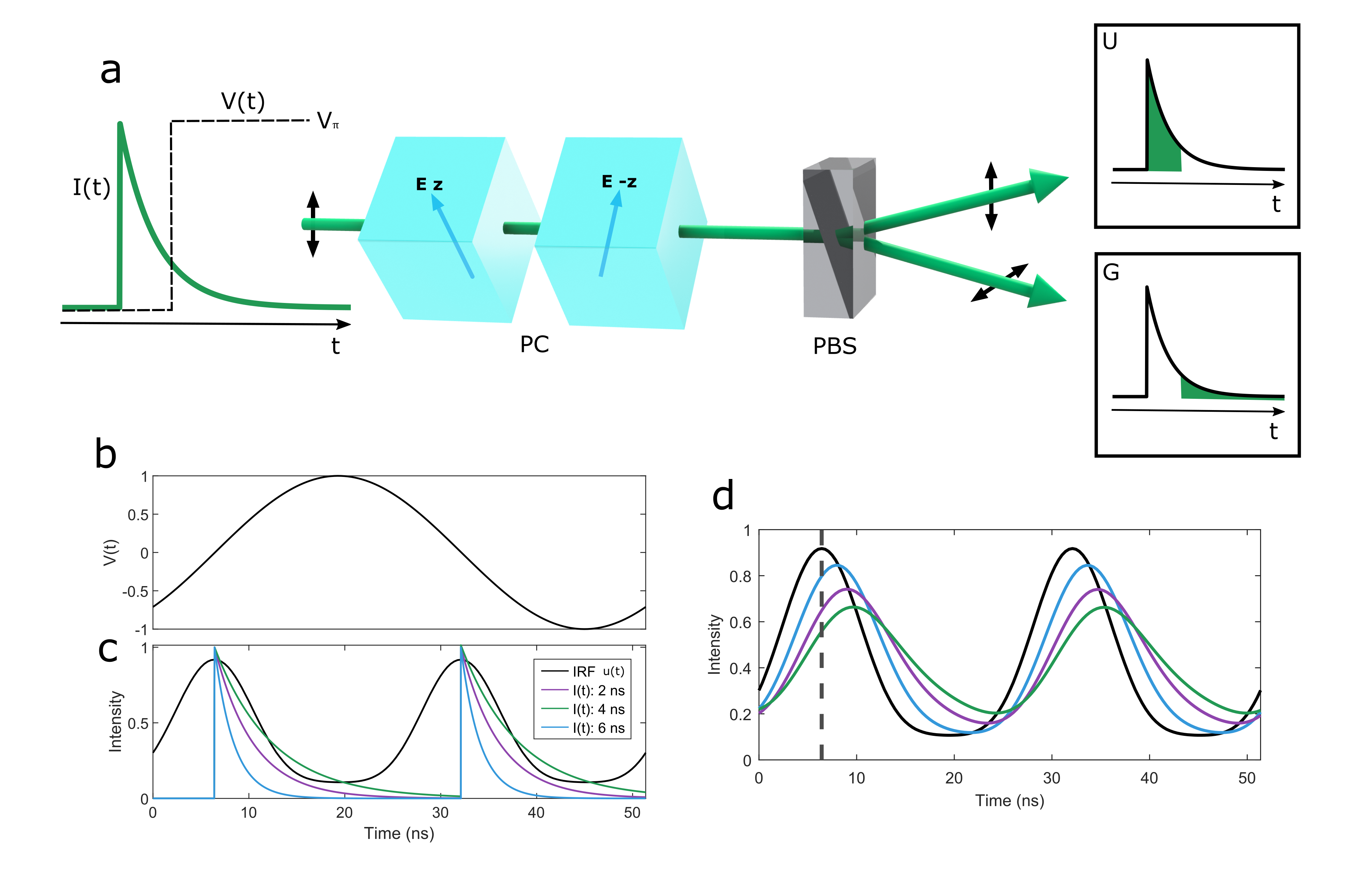} 
  \vspace{-5.0mm}
\caption{\label{fig:1} {EO-FLIM method. \textbf{(A)} Schematic diagram of EO-FLIM gating. A fluorescence decay at left is gated by an ideal voltage step function applied to the Pockels cell (PC) at a time delay after the excitation pulse. Light arriving before the voltage step is captured in ungated output image (U) after the polarizing beamsplitter (PBS). Light arriving after the voltage step has its polarization rotated by 90 degrees and is captured in gated output image (G). All light from the original decay is preserved up to optical losses. To optimize imaging performance, a transverse field lithium tantalate Pockels cell is employed with electric field applied perpendicular to the optical axis. In order to compensate for static and off-axis birefringence, a pair of crystals is used with a 90 degree relative rotation (crysal Z axis indicated) and reversed electric field directions. \textbf{(B)} Instead of an ideal step-function, our experiments applied a sinusoidal voltage drive to the crystal with $\sim$1.2 kV amplitude. \textbf{(C)} Experimentally realized gating function u(t) in the ungated channel is plotted along with simulated 2, 4, and 6 ns fluorescence decays, I(t). This relative phase setting is used for all acquired data. \textbf{(D)} Plot of output intensity U as a function of phase delay between excitation laser pulses and u(t). This represents the convolution of I(t) with u(t). Simulated phase traces are plotted for the three lifetimes in (C). For fast voltage imaging, a single phase point from this trace is sampled, corresponding to one camera exposure (vertical line) where lifetime is estimated from the ratio of G and U image intensities. Note that $u(t)+g(t)=1$, so g(t) is not shown here. 
}}
\end{figure*}

\newpage
\begin{figure*}[h!]
\centering
  \vspace{-15.0mm}
  \includegraphics[width=0.8\textwidth]{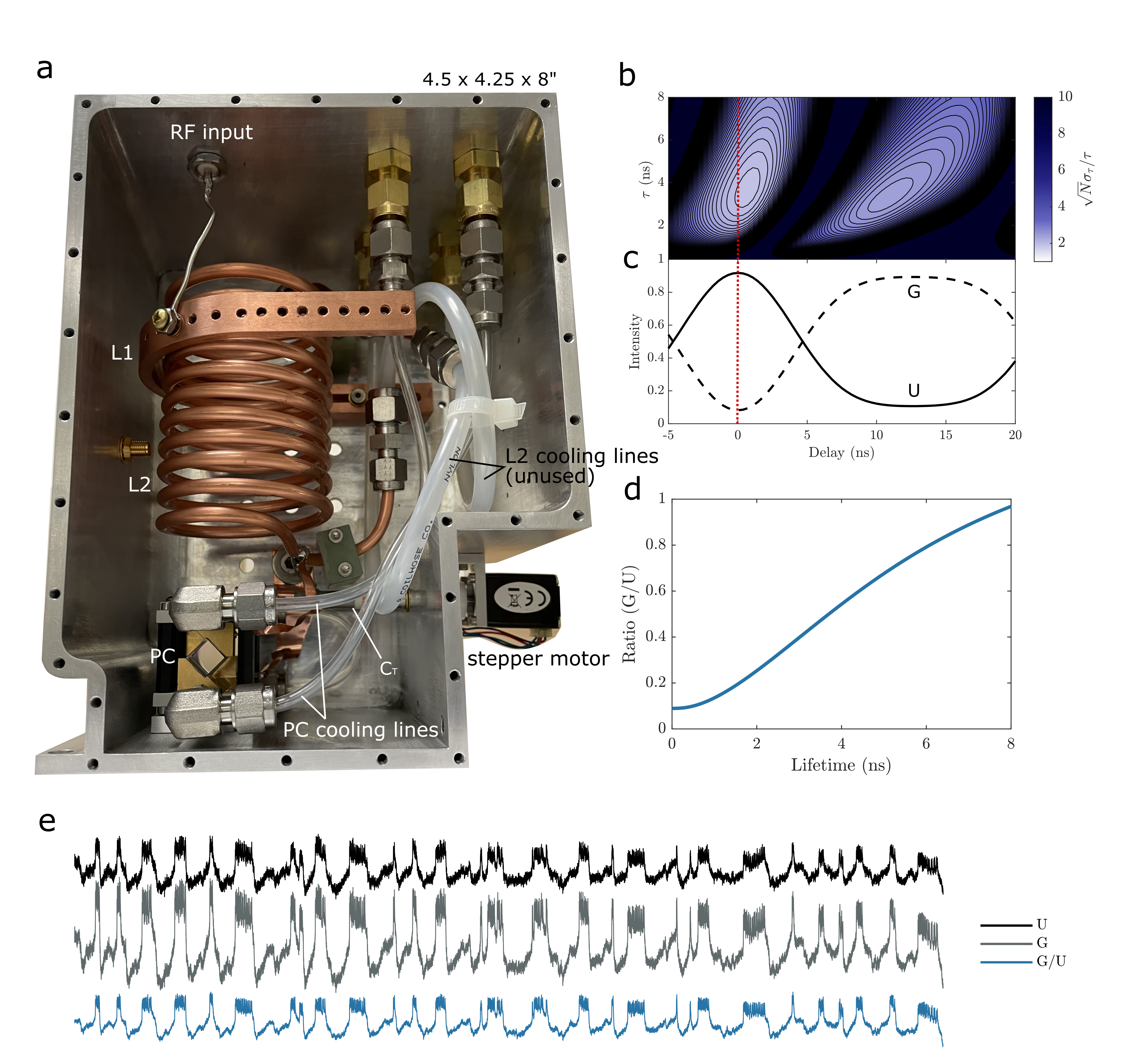} 
  \vspace{-5.0mm}
\caption{\label{fig:1} {Resonant Pockels cell gating. (\textbf{A}) Pockels cell (PC - aperture at bottom left) and radio frequency (RF) drive circuit inside grounded enclosure. A resonant transformer is used to present a high-voltage drive to the Pockels cell using 9.3 W of input RF power. The transformer is tuned to provide a 50 ohm load to the driving electronics. The primary L1 is a half turn copper strap, and the secondary L2 consists of a 11 turn coil of copper tubing connected in series with the Pockels cell, which acts as a capacitor. A trim capacitor $\mathrm{C_T}$ is adjusted by a stepper motor to tune the circuit's resonance frequency. Water cooling is provided to the PC, and additional lines for oil cooling of L2 were included but not used during normal operation. The entire assembly is mounted on a tilt and rotation platform for alignment (Newport model 36). (\textbf{B}) Photon normalized estimation accuracy (F number or $\sqrt N\sigma_\tau/\tau$) for EO-FLIM is plotted as a function of lifetime and phase delay using the experimental instrument response function plotted in (C). F=1 corresponds to the fundamental limit for an optimal lifetime measurement. (\textbf{C}) Instrument response function measured with quenched fluorescein dye. The gated (G) and ungated (U) image intensities are normalized and plotted as a function of phase delay as in Fig. S1. (\textbf{D}) Lookup table used for converting a measured intensity ratio G/U to a fluorescence lifetime value. In the absence of experimental noise, lifetime error is determined by photon shot noise in the G and U channels and the slope of the lookup table function. (\textbf{E}) Example voltage imaging traces for gated (G), ungated (U), and ratio (R) channels are plotted on the same vertical scale normalized to their mean values. The direct G and U traces show that despite $80\%$ of intensity being captured in U, the G channel has enhanced spike height and voltage responsivity. This happens because a positive intensity change also results in a positive lifetime change, which transfers intensity from U into G. The ratio signal shows attenuated overall response but improved SNR. It is directly proportional to lifetime for small changes since the lookup table in (D) is locally linear.
}}
\end{figure*}

\newpage
\begin{figure*}[h!]
\centering
  \includegraphics[width=1.0\textwidth]{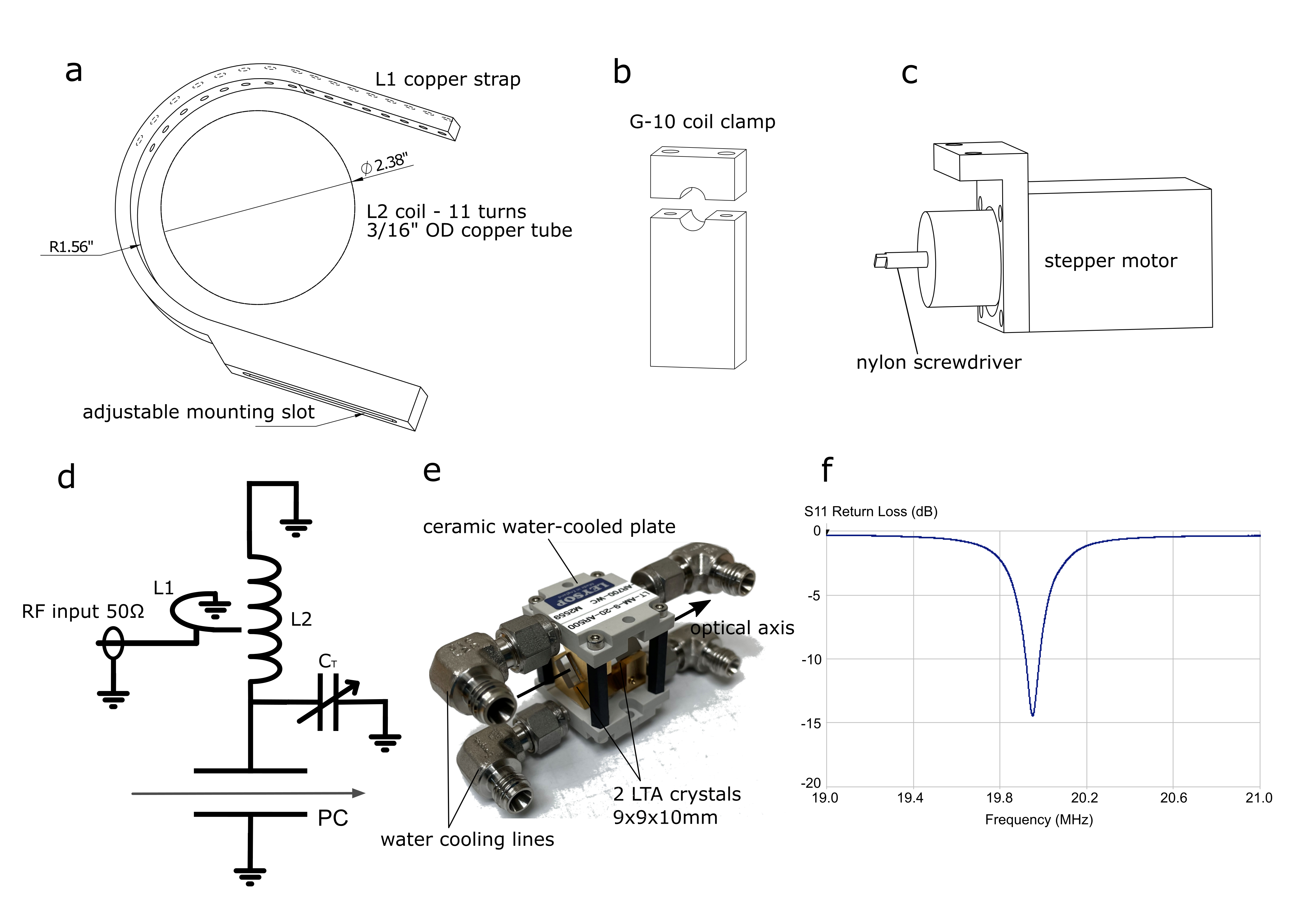} 
  \vspace{-5.0mm}
\caption{\label{fig:1} {System components. \textbf{(A)} Mechanical drawing of primary (L1) and secondary (L2) inductor dimensions. L2 is wound from 11 turns of 3/16" OD copper refrigeration tubing. While provisions for oil cooling were provided in the design, it was not necessary to cool L2 during operation. Adjustable tap points and mounting slot allow impedance matching to 50 $\Omega$ \textbf{(B)} Fiberglass (G-10) supports clamp  the L2 coil into place \textbf{(C)} A stepper motor and screwdriver adapter allow remote tuning of a 10 pF high voltage trim capacitor $\mathrm{C_T}$ (Sprague-Goodman SGNMNC1103) to adjust the circuit's resonant frequency. \textbf{(D)} Electrical schematic of the resonant transformer. The primary L1 is tuned to present a 50 $\Omega$ load to the drive amplifier. \textbf{(E)} Pockels cell assembly with two lithium tantalate crystals (LTA) and optical axis indicated. The Pockels cell electrodes are water cooled through thermally conductive but electrically insulating aluminum nitride ceramic plates. \textbf{(F)} Plot of S11 return loss for the resonant transformer. An unloaded Q factor of 117 is measured which is significantly reduced under RF drive. With 9.3W input (31 V into 50 $\Omega$) a $\sim$1.2 kV drive is achieved across the Pockels cell.
}}
\end{figure*}

\newpage
\begin{figure*}[h!]
\centering
  \includegraphics[width=0.5\textwidth]{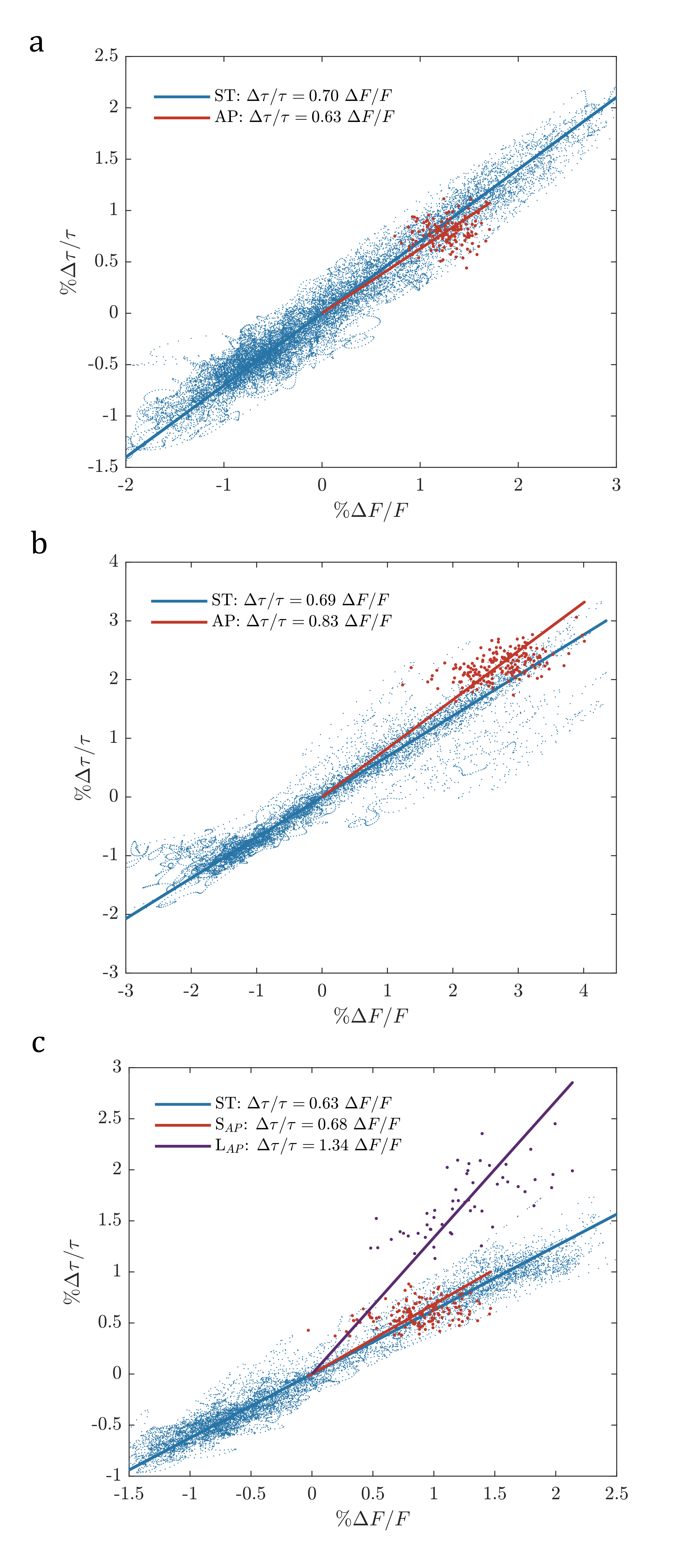} 
  \vspace{-5.0mm}
\caption{\label{fig:1} {Lifetime response linearity. Scatter plots of $\Delta F/F$ and $\Delta \tau/\tau$ are shown with panels A-C corresponding to the traces in Figs. 1(E), 2(A), and 2(F) respectively.  Sub-threshold waveforms are plotted with a 20 Hz low-pass filter applied (blue), while action potentials are plotted separately (red) to indicate high-frequency response. Both estimates provide comparable responsivity relationships. Combined with existing calibrations for $\Delta F/F$, these plots imply a linear relationship between lifetime and membrane potential (\textit{3, 7}). In (C), the L spikes from Fig. 2(F) are observed to have much larger lifetime responsivity than the S spikes.
}}
\end{figure*}

\newpage
\begin{figure*}[h!]
\centering
  \includegraphics[width=0.9\textwidth]{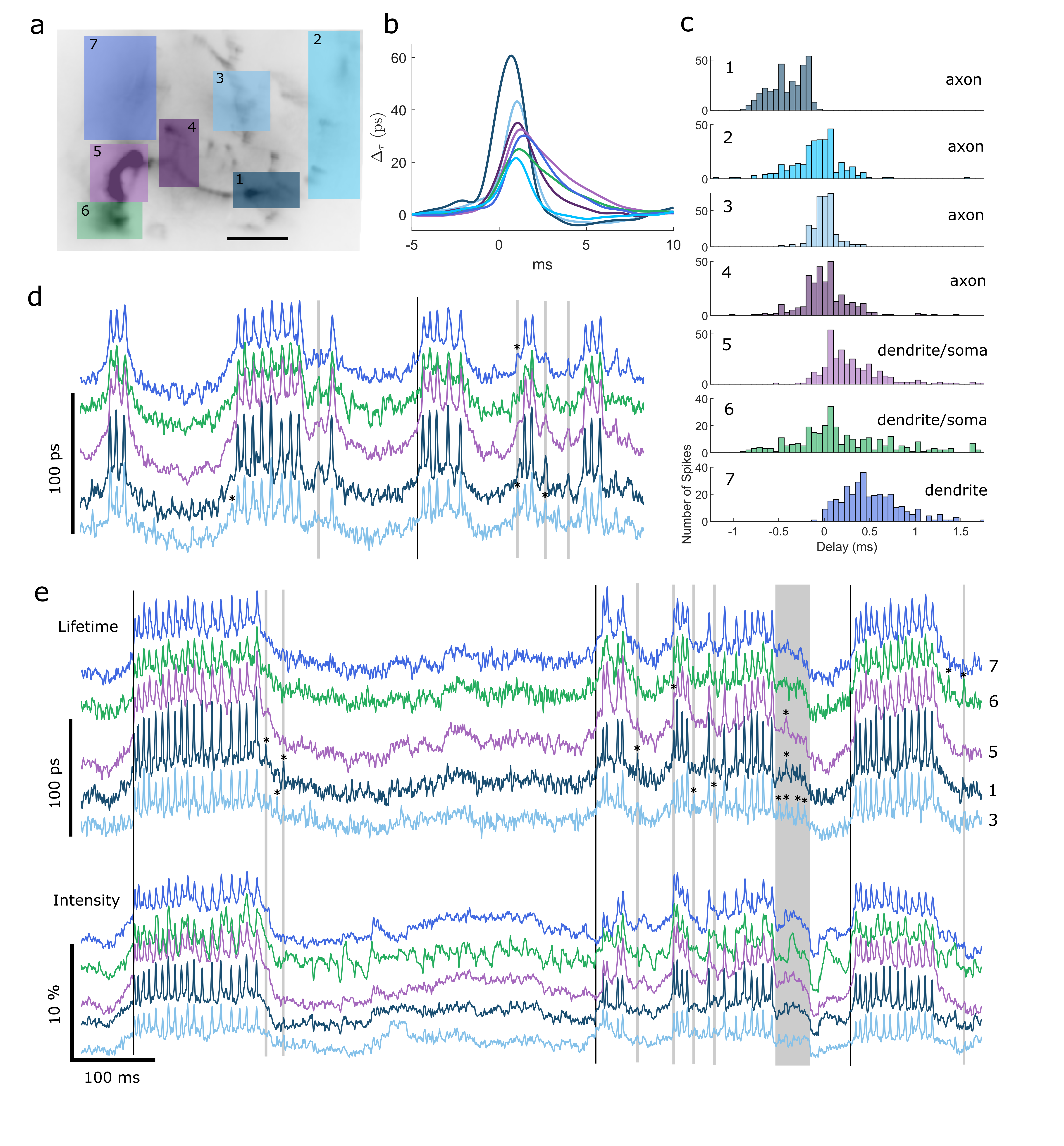} 
  \vspace{-5.0mm}
\caption{\label{fig:1} {Spike propagation and ROI analysis. \textbf{(A)} ROIs are defined for several labeled neuron sub-regions. \textbf{(B)} Average spike waveforms are calculated from spline-interpolated traces. \textbf{(C)} Spike delay histograms are shown. Times are calculated using the peaks of the interpolated waveform relative to the average spike time for the whole structure. Bidirectional propagation from the point of initiation is observed both to more distant axons and backwards towards dendrites and soma. \textbf{(D,E)} Comparing lifetime and intensity ROI traces (three frame moving averages) reveals stronger sub-threshold responses in the dendrites. Local effects can also be observed. Notably small spike-like depolarizations occur in the axons which do not trigger action potentials (indicated with asterisks). These depolarizations are often not resolved in the intensity trace.
}}
\end{figure*}

\newpage
\begin{figure*}[h!]
\centering
  \includegraphics[width=0.9\textwidth]{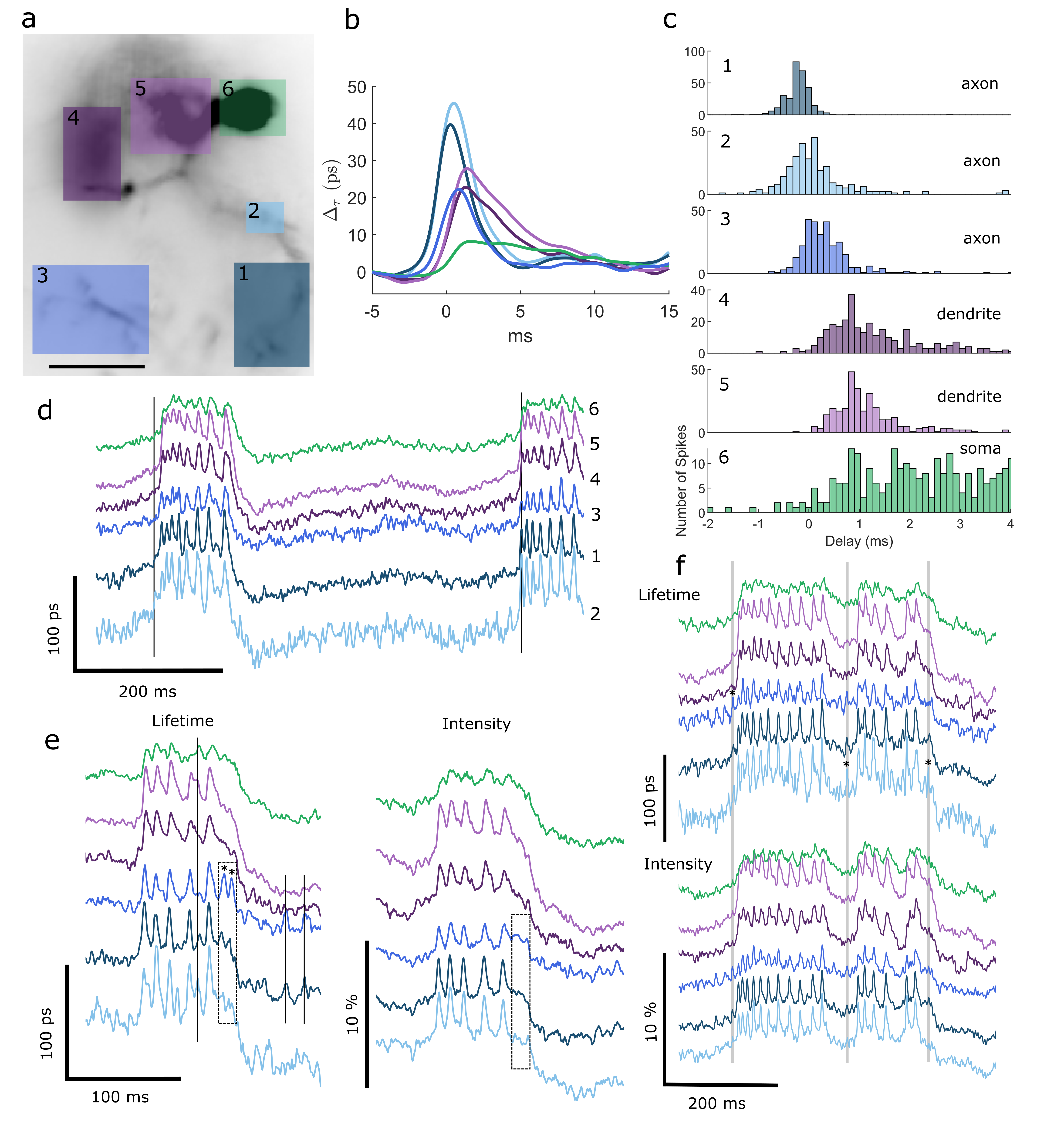} 
  \vspace{-5.0mm}
\caption{\label{fig:1} {Spike propagation and ROI analysis. \textbf{(A)} ROIs are defined for several labeled neuron sub-regions. \textbf{(B)} Average spike waveforms are calculated from spline-interpolated traces. \textbf{(C)} Spike delay histograms are shown. Times are calculated using the peaks of the interpolated waveform relative to the average spike time for the whole structure. Bidirectional propagation from the point of initiation is observed both to more distant axons and backwards towards dendrites and soma. \textbf{(D-F)} Comparing lifetime and intensity ROI traces (three frame moving averages) reveals stronger sub-threshold responses in the dendrites. Local effects can also be observed. Notably small spike-like depolarizations occur in the axons which do not trigger action potentials (indicated with asterisks). These depolarizations are often not resolved in the intensity trace.
}}
\end{figure*}

\newpage
\begin{figure*}[h!]
\centering
  \includegraphics[width=0.9\textwidth]{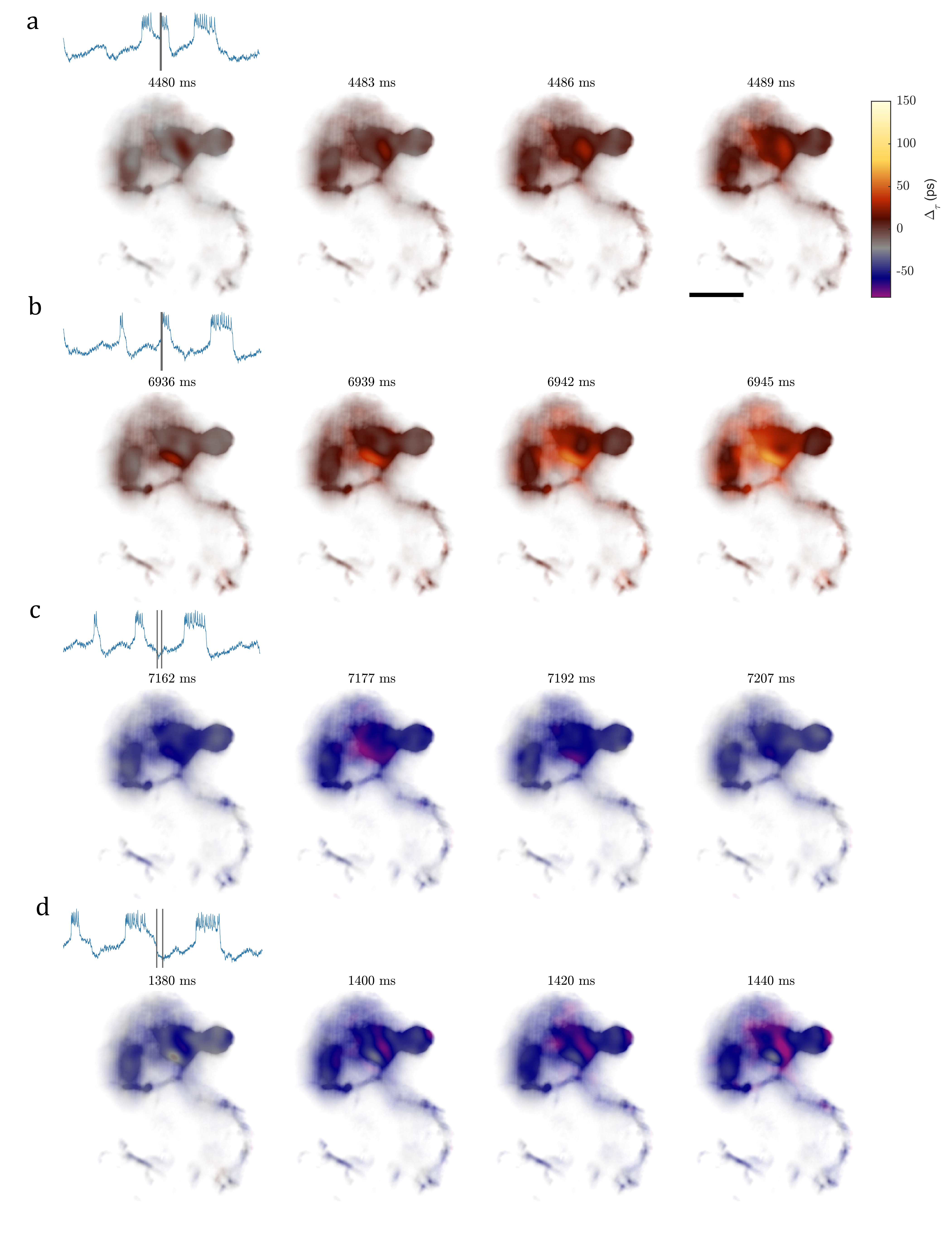} 
  \vspace{-5.0mm}
\caption{\label{fig:1} {Sub-threshold voltage signals localize in the dendrite. Four time-lapse image series are displayed from Movie S6 with corresponding time points indicated on the inset traces. (\textbf{A},\textbf{B}) Some positive-going transitions appear to originate in localized regions of the dendrite. Negative sub-threshold features (\textbf{C},\textbf{D}) have slower time dynamics but also show increased strength in the dendrite (see also Fig. S6 and Movies S6 and S7). In Movie S6 a 10 frame moving average was applied to the raw image data (scalebar 25 $\mu$m).
}}
\end{figure*}

\newpage
\begin{figure*}[h!]
\centering
  \includegraphics[width=0.9\textwidth]{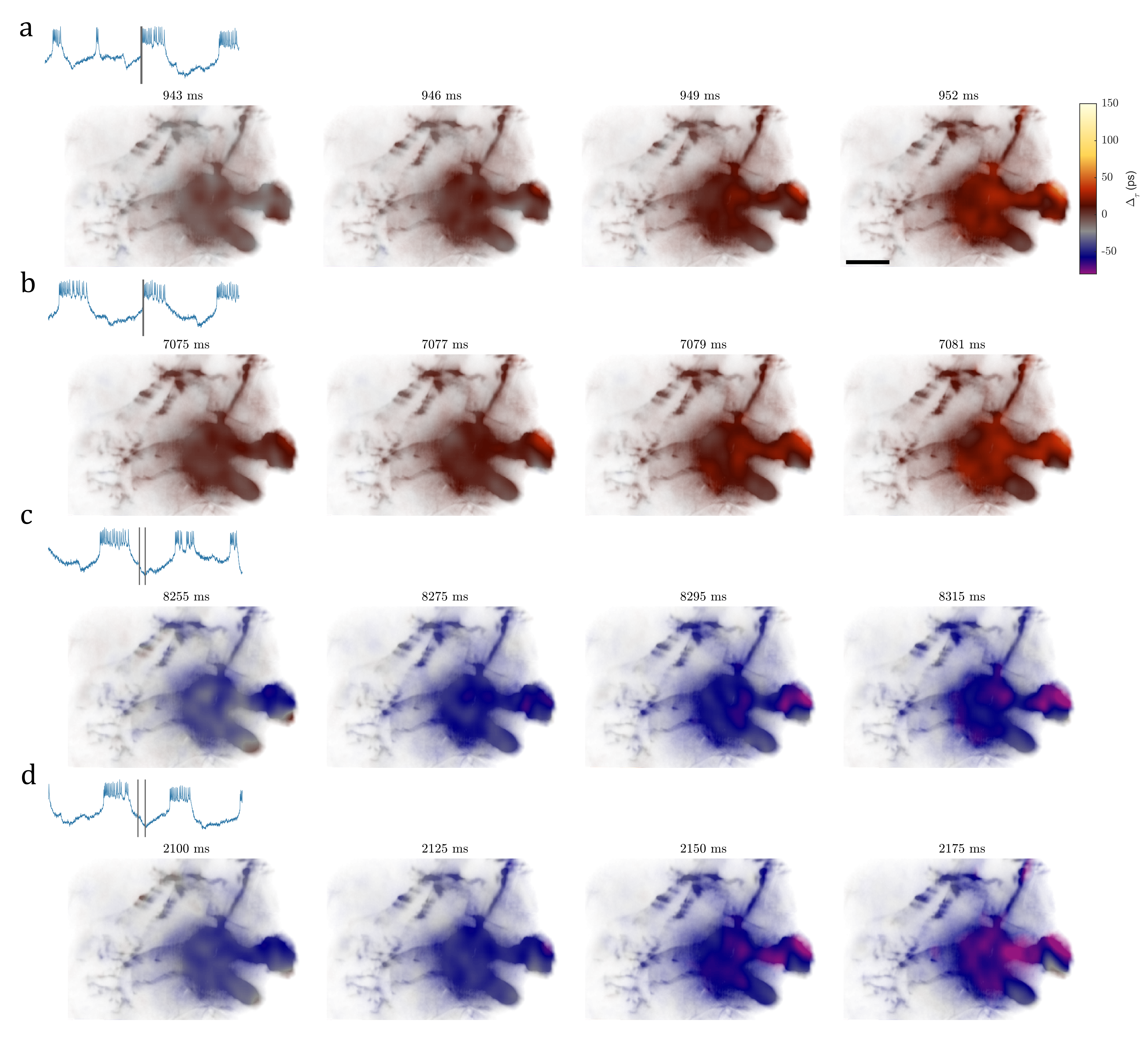} 
  \vspace{-5.0mm}
\caption{\label{fig:1} {Sub-threshold voltage signals localize in the dendrite. Four time-lapse image series are displayed from Movie S7 with corresponding time points indicated on the inset traces. Positive (\textbf{A},\textbf{B}) and negative (\textbf{C},\textbf{D}) sub-threshold voltage changes show increased strength in the dendrite. In Movie S7 a 10 frame moving average was applied to the raw image data (scalebar 25 $\mu$m).
}}
\end{figure*}

\newpage
\begin{figure*}[h!]
\centering
  \includegraphics[width=0.9\textwidth]{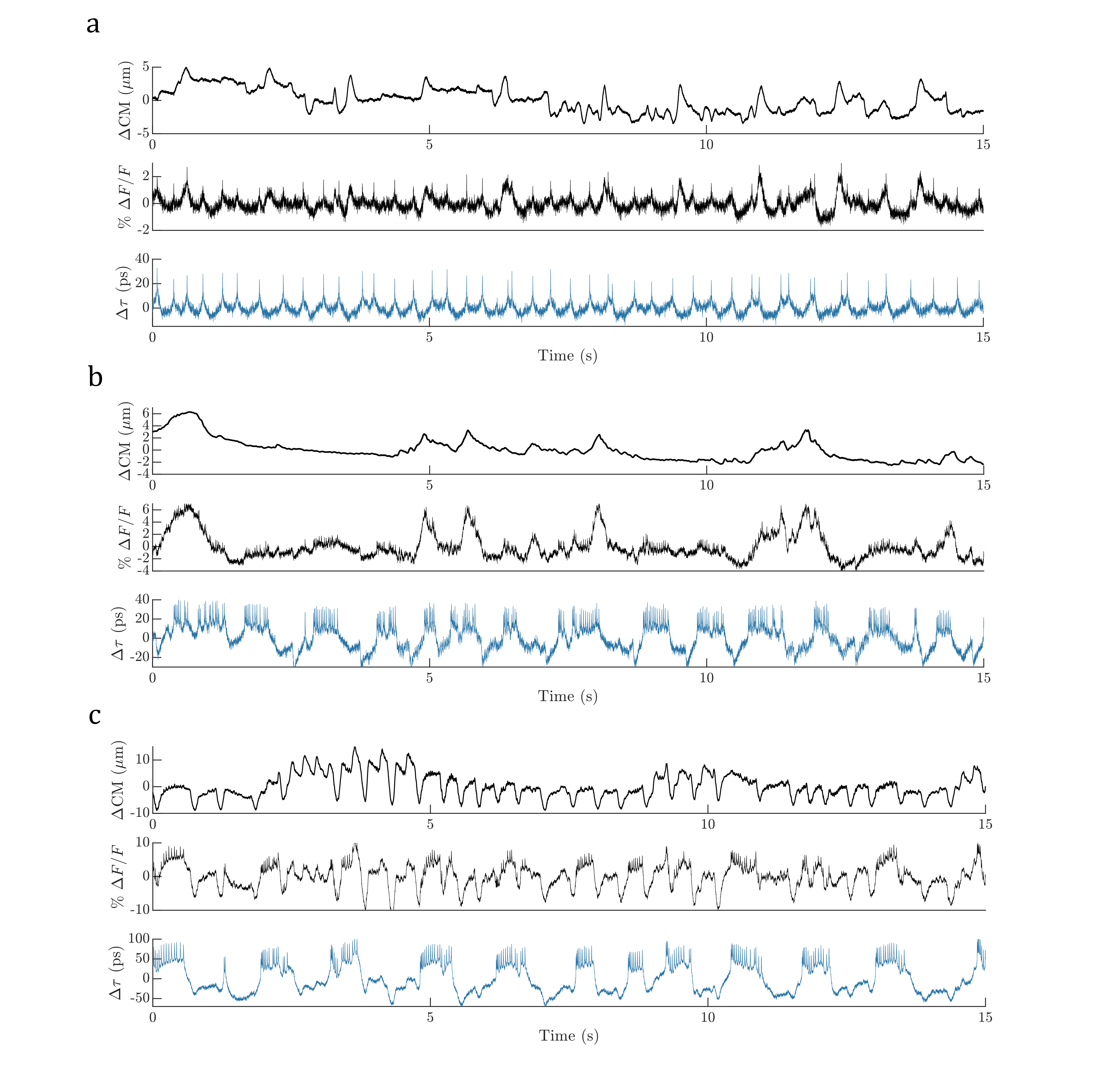} 
  \vspace{-5.0mm}
\caption{\label{fig:1} {Motion artifacts. Center of mass motion traces and lifetime and intensity traces are given in (A-C) for the data presented in Fig. 2(C-E). Areas with intensity artifacts strongly correlate with sample motion. Motion in the Z plane is only partially captured here. 
}}
\end{figure*}

\newpage
\begin{figure*}[h!]
\centering
\hspace*{-0.5in}
  \includegraphics[width=1.15\textwidth]{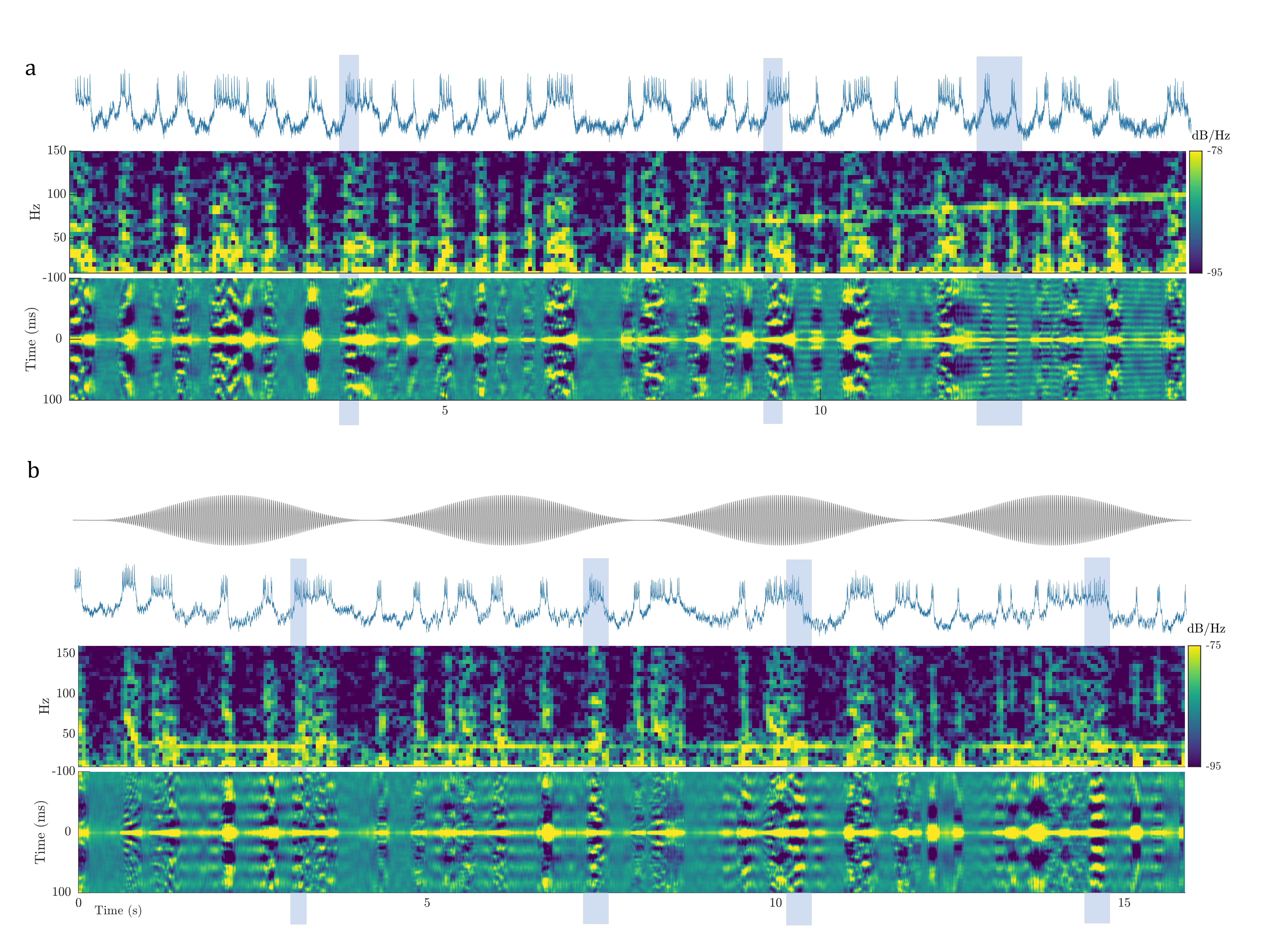} 
  \vspace{-5.0mm}
\caption{\label{fig:1} {Two additional examples of mechanically evoked phase locking. (\textbf{A}) Direct XY shaking stimulus ramp similar to Fig. 4. (\textbf{B}) An example using audio stimulation with periodic envelope drive (waveform displayed at top). We note that the start and stop of mechanical stimulus is usually correlated with a spike activity burst. A few example regions with phase locking are highlighted with shaded boxes. Some spike bursts demonstrate transitions between unlocked and locked behavior.
}}
\end{figure*}

\newpage
\begin{figure*}[h!]
\centering
  \includegraphics[width=0.70\textwidth]{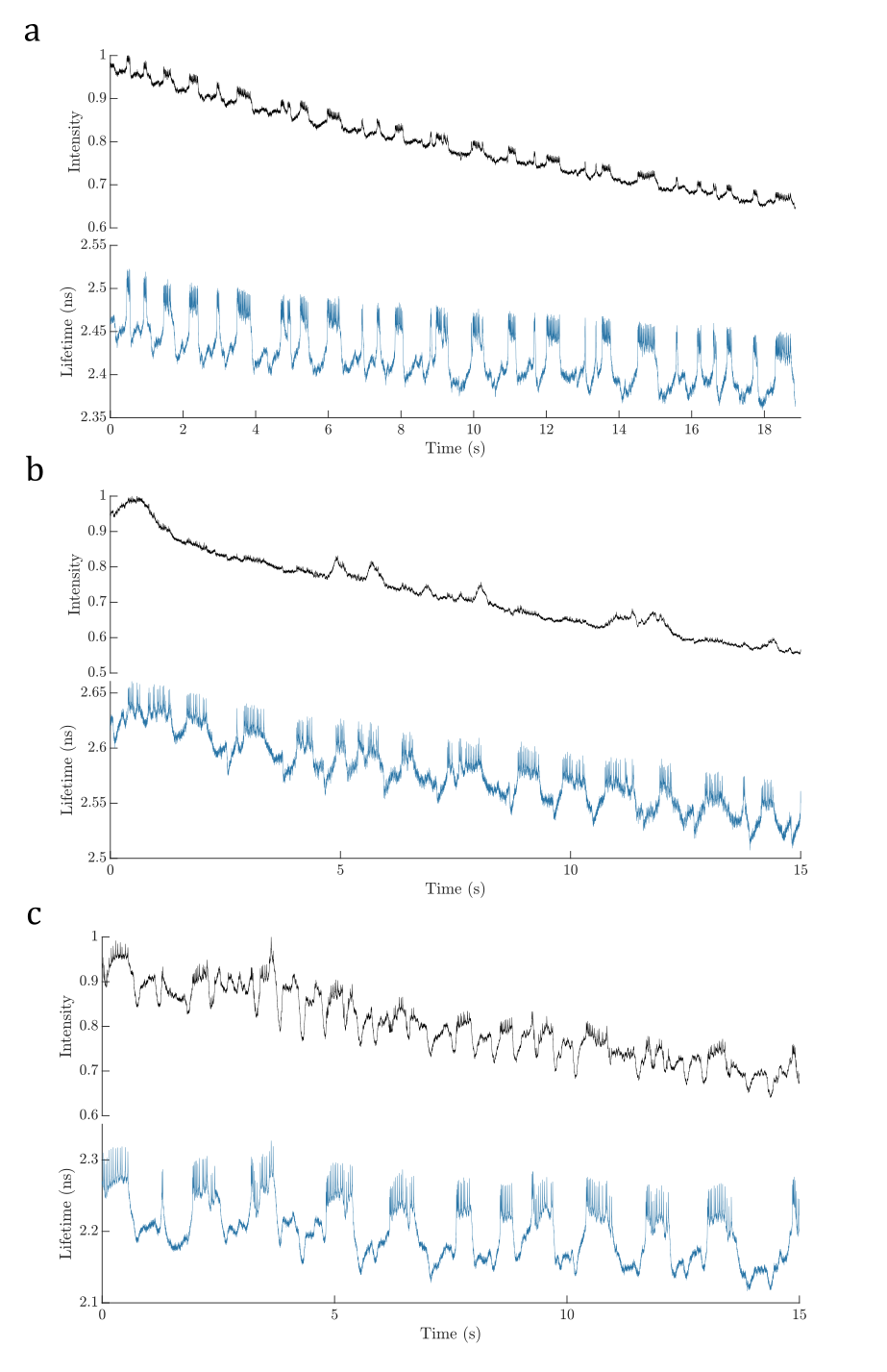} 
  \vspace{-5.0mm}
\caption{\label{fig:1} {Unprocessed intensity and lifetime traces show that the effects of photobleaching on GEVI intensity readout (up to 30$\%$ during a recording) are strongly suppressed by lifetime readout. Due to the presence of non-specific background fluorescence with different bleaching rate, however, there are small lifetime offsets from sample to sample and slow drifts in lifetime at the 50-100 ps level. Here plots A,B, and C correspond to the data in Figs. 1(E),  2(D), and 2(E) respectively.
}}
\end{figure*}

\FloatBarrier
\newpage
\noindent\textbf{Movie S1.} Spike trigger averaged movie of action potential propagation (scalebar 25 $\mu$m).\\\\
\textbf{Movie S2.} Spike trigger averaged movie of action potential propagation (scalebar 25 $\mu$m).\\\\
\textbf{Movie S3.} Spike trigger averaged movie of action potential propagation (scalebar 25 $\mu$m).\\\\
\textbf{Movie S4.} Non-averaged movie of action potentials (scalebar 25 $\mu$m).\\\\
\textbf{Movie S5.} Non-averaged movie of action potentials (scalebar 25 $\mu$m).\\\\
\textbf{Movie S6.} Movie of sub-threshold voltage activity (10 frame moving average applied, scalebar 25 $\mu$m).\\\\
\textbf{Movie S7.} Movie of sub-threshold voltage activity (10 frame moving average applied, scalebar 25 $\mu$m).\\\\
\textbf{Movie S8.} Spike trigger averaged movie of small spikes from Fig. 2(F-L) (scalebar 25 $\mu$m).\\\\
\textbf{Movie S9.} Spike trigger averaged movie of large spikes from Fig. 2(F-L) (scalebar 25 $\mu$m).\\\\

\end{document}